
\baselineskip=14pt
\magnification=\magstep1%\def\DRAFT{1}

%\parskip=\smallskipamount 

\def\quote#1{\medskip
{\baselineskip=12pt{\narrower\narrower\parindent = 0pt #1
\par}\medskip}}

\def\abst#1{\medskip{\baselineskip=12pt
{\narrower\narrower\parindent = 0pt #1 \par}}} 

\newcount\ftnumber
\def\ft#1{\global\advance\ftnumber by 1
          {\baselineskip=12pt \footnote{$^{\the\ftnumber}$}{#1 \par}}}

\newcount\fnnumber
\def\fn{\global\advance\fnnumber by 1
         $^{(\the\fnnumber)}$}

\def\fr#1/#2{{\textstyle{#1\over#2}}} 

\def\>{\rangle}
\def\<{\langle}
\def\k#1{|#1\>}
\def\b#1{\<#1|}
\def \ip#1#2{\< #1 | #2 \>}
\def\x{\otimes}
\def\X{\otimes}
\def\s{{\cal S}}

\def\tr{{\rm Tr}}
\def\nv{{\bf n}}
\def\xv{{\bf x}}
\def\yv{{\bf y}}
\def\zv{{\bf z}}
\def\u{\uparrow}
\def\d{\downarrow}
\def\ua{\u_a}
\def\da{\d_a}
\def\ub{\u_b}
\def\db{\d_b}

\def\uc#1{\u_{c#1}}
\def\dc#1{\d_{c#1}}
\def\c{\fr1/{\sqrt2}}
\def\oo{\overline{1}}
\def\al{\alpha}

\def\x{\otimes}
\def\h{\fr1/2}

\def\D{\bar D}

\newcount\eqnumber

\def\eq(#1){
    \ifx\DRAFT\undefined\def\DRAFT{0}\fi	%if undef'd, make it 0
    \global\advance\eqnumber by 1%
    \expandafter\xdef\csname !#1\endcsname{\the\eqnumber}%
    \ifnum\number\DRAFT>0%
	\setbox0=\hbox{#1}%
	\wd0=0pt%
	\eqno({\offinterlineskip
	  \vtop{\hbox{\the\eqnumber}\vskip1.5pt\box0}})%
    \else%
	\eqno(\the\eqnumber)%
    \fi%
}
\def\(#1){(\csname !#1\endcsname)}

%\rightline{ \number\month /\number\day /\number\year}
%; (\number\time)} 

\centerline{{\bf What Do These Correlations Know about Reality?}}
\centerline{{\bf Nonlocality and the Absurd\/}\ft{{Dedicated to Daniel
Greenberger on the occasion of his 65th birthday. A version will
appear in Foundations of Physics.}}}

\bigskip
\centerline{N. David Mermin}
\centerline{Laboratory of Atomic and Solid State Physics}
\centerline{Cornell University, Ithaca, NY 14853-2501}
\bigskip
\abst{In honor of Daniel Greenberger's 65th birthday I record for
posterity two superb examples of his wit, offer a proof of an
important theorem on quantum correlations that even those of us over
60 can understand, and suggest, by trying to make it look silly,
that invoking ``quantum nonlocality'' as an explanation for such
correlations may be too cheap a way out of the dilemma they pose.} 

\bigskip
\noindent{{\bf 1. INTRODUCTION}}
\medskip

Daniel Greenberger has a fine sense of the absurd.  He can be
breath-takingly funny at the most unexpected moments, leaving you
gasping in admiration.  I consider it my duty to record
for the pleasure of future generations two examples of his wit in
which I am privileged to have served as straight man.

My first example tells us something about Danny.  I submitted a
paper\fn\ explaining how the three particle version of the
Greenberger-Horne-Zeilinger argument can be reformulated as an
extremely simple version of the more complicated Bell-Kochen-Specker
theorem.  Dimly remembering Danny long ago making incomprehensible
remarks about why it might be interesting to think about a particle
decaying into a pair of particles each of which then decayed into a
pair of its own, I carelessly began a footnote with ``Greenberger invented his
example to$\,\ldots\,$.'' In due course the report of an
anonymous referee arrived.  Here it is in full:

\medskip
\quote{A very pretty paper.  Unfortunately it is rendered unpublishable by
virtue of the mischievous footnote 19, which is an affront to what is
left of Western Civilization.  When three people agree to work
together on a project, it is very unfair to single out one for most of
the credit.  Regrettably, because physicists are hero-worshippers,
they often do this anyway, especially if one of the group is very
famous.  Thus, everyone usually mentions only Einstein in connection
with EPR.}
\quote{But to accord such treatment to a bonehead like Greenberger is
unforgivable.  I recommend the paper be sent back to the author for
correction (``The GHZ example was invented to$\ldots$'' or better, ``The lovely
GHZ example was invented to$\ldots$'') and upon resubmission, it should be
published without further refereeing or delay.}

My second example, I rashly suggest, serves as a metaphor for how
not to think about quantum nonlocality.  After a conference on
foundations of quantum mechanics in Urbino about a dozen years ago,
Danny and I spent a couple of days travelling around Tuscany and
Umbria.  The moment of the creation of the finest spontaneous joke in
human history finds us in the Academy in Florence, standing in front
of Michaelangelo's David:

\quote{{\it Mermin:} Look, he's not circumcised.}
\quote{{\it Greenberger} [instantly]: Ah, what do these {\it
goyim\/} know about sculpture? } I would like to suggest that the same
kind of exuberantly setting sail in entirely the wrong direction takes
place when one is led to infer physical nonlocality from the strange
correlations in EPR or GHZ or Hardy states.  The straight response to
my remark about David's puzzling anatomy would have been ``What do
these {\it goyim\/} know about Jewish ritual?'' That lunatic last minute
switch from Levitical legislation to Michaelangelo's qualifications as
an artist is analogous to the breathtaking switch we make from the
inability of physics to deal with the individual properties of an
individual system, to the miraculous creation of such properties from
afar. 

This, of course, is just a hunch.  I will not establish it here by
direct argumentation, but will try at least to undermine the notion
that nonlocality is a sensible inference from quantum entanglement, by
a method which, inspired by Danny,\ft{Inventor --- no, I mean
co-inventor --- of that essential guide to clear thinking, the Law of
the Excluded Muddle.} I make bold to call a process of successive
exasperations.  I give the main {\it reductio ad absurdum\/} in
Section 3. I describe in Section 2 some attitudes toward
non-locality, give a preliminary, familiar, run through the successive
exasperations, and describe a pertinent theorem (whose elementary
proof is in the Appendix) which deserves to be more widely known.  

\bigskip
\noindent{\bf 2. AGAINST NONLOCALITY: THE OBJECTIVE NATURE OF
MIXED STATES}
\medskip 

Sir Rudolf Peierls didn't believe Bell's theorem established
nonlocality.  It only showed that any attempt to complete quantum
mechanics with hidden variables would be necessarily
nonlocal.\fn\ On one of his visits to
Ithaca I put it to him that he must therefore take the view that the
real lesson of EPR-Bell is not that there is objective nonlocality,
but that there is no objective difference between a 50-50 ensemble of
horizontally ($90^\circ$, $H$) and vertically ($0^\circ$, $V$)
polarized photons, and a 50-50 ensemble of diagonally ($\pm 45^\circ$,
$D$ and $\bar D$) polarized photons.  He said he did.  

The view that different realizations of one and the same mixed-state
density matrix describe objectively identical ensembles is one of
those propositions whose validity can flicker on and off like an
optical illusion.  On the one hand there is obviously no objective
difference between a 50-50 mixture of $H$ and $V$ polarizations and a
50-50 mixture of $D$ and $\bar D$, since the statistics of any
measurements you can make on each of the two ensembles are identical.  

On the other hand, there obviously {\it is\/} an objective difference,
since the photons in the two ensembles have different histories,
having been prepared in different ways.  In one case Bob can give
the ensemble to Alice by sending her a series of photons each of which
has emerged from a polarizer randomly chosen each time to transmit
either horizontal or vertical polarizations; in the second case his
polarizer randomly transmits either of the two diagonal
polarizations.  So the photons have quite distinct past experiences in
the two cases, which makes the two ensembles objectively different.
Furthermore Bob can demonstrate the difference to Alice by telling
her, in the first case, the results she would get if she were to measure
the horizontal-vertical polarization of each photon, or, in the second
case, the results she would get for their diagonal polarizations 

But on further reflection, there obviously {\it is not\/} an objective
difference since Bob can also do this trick by preparing pairs of
photons in the polarization state $\fr1/{\sqrt2}(\k{H,H} + \k{V,V}) =
\fr1/{\sqrt2}(\k{D,D} + \k{\bar D,\bar D})$, and sending one member of
each pair to Alice.  He can then get the information he needs to inform
Alice of the results she will get for either set of experiments by
measuring the appropriate polarization of the member of
each pair that he keeps for himself.   --- i.e. without doing anything
whatever to Alice's photons, whose past experiences are identical in the
two cases.  

Yet on subsequent reconsideration, if one believes in quantum
nonlocality then, after all, there still {\it is\/} an objective
difference, for by subjecting his member of each pair to a polarization
measurement Bob is, in fact, altering the character of Alice's
corresponding member in a way that depends on his choice of which
polarization to measure.  The alteration takes place through the well
known mechanism of spooky action at a distance, which might better be
called the {\it SF\/} mechanism, in recognition of Einstein's original
term, {\it spukhafte Fernwirkungen}.\fn\

Even so, however, if you look a little more carefully, then, in the final
analysis there clearly {\it is not\/} an objective difference,
since Alice can measure either the horizontal-vertical or diagonal
polarizations of all her photons, indelibly recording the data in a
notebook, long before Bob ever starts to measure any of his corresponding
photons.  Bob's acts of measurement cannot possibly alter the
character of the data in Alice's notebook, nor can they alter
the character of her photons, which can be destroyed in the process of
her polarization measurements long before Bob attempts to alter them
via the {\it SF\/} mechanism. 

This last flip-flop strikes me as definitive.  While it may be
reasonable to contemplate the possibility that Bob's measurement can
alter the character of something as wraith-like as the potentialities of
Alice's photons for subsequently behaving in one way or another, if the
behavior has already taken place and been recorded in ink before Bob
takes any action, then nobody\ft{Except those who think that it's
looking in the box that kills Schr\"odinger's cat.} would maintain
that the {\it SF\/} mechanism can affect the notebook entries.  

And indeed, if you believe in {\it SF\/}, then the least implausible
thing to say by way of refutation is that if Alice makes her
measurements long before Bob makes his, then she is at that very time
inadvertently and spookily transmitting to him the very information he
will later need to {\it deceive her\/} into thinking that he prepared the
ensemble, which she herself disposed of some time before, as either a
horizontal-vertical or a diagonal ensemble.  At this point {\it SF\/}
takes on the character of a clever (if not uproariously funny) joke.

I tell in Section 3 a somewhat more subtle (still --- sorry, Danny ---
not rib-tickling) version of this joke, which, at least for me, makes
{\it SF\/} look even more like science fiction.  

\medskip

This apparent ability remotely to produce distinct ensembles with the
same density matrix $W$ might appear to stem from the degeneracy of
$W$, but in fact the trick\ft{I use the word ``trick'', because the
situation is reminiscent of stage magic, which works by a process of
self-deception, skillfully helped along by the magician.  When the
trick is {\it SF\/}, the magician is Nature herself.} can be done quite
generally.  Take as a simple example the non-degenerate density matrix
$$W = p\k{H}\b{H} + q\k{V}\b{V}\eq(3.5)$$ with $p \neq q$, which in
spite of its non-degeneracy has many alternative representations in
terms of {\it nonorthogonal\/} polarization states, for example $$W =
\h\k{R}\b{R} + \h\k{L}\b{L},\eq(3.6)$$ where and $\k R$ and $\k L$ are
$$\k R = \sqrt p\k{H} + \sqrt q\k{V}\eq(3.7)$$ and $$\k L =
\sqrt p\k{H} - \sqrt q\k{V}.\eq(3.8)$$ Representation \(3.5) of $W$
describes an ensemble of photons that are horizontally polarized with
probability $p$ and vertically polarized with probability $q$;
representation \(3.6) describes an ensemble of photons that have with
equal probability one of two linear polarizations along axes tilted
away from \zv\ along \xv\ or $-\xv$ through an angle $\theta =
\cos^{-1}(\sqrt p)$.  

One can produce either one of these ensembles from afar, in the EPR
manner, by introducing a second non-interacting two-state system and
starting with an ensemble of pairs, each in the pair state $$\k\Psi =
\sqrt p\k{H}\X\k{H} + \sqrt q \k{V}\X\k{V}. \eq(3.9)$$ This state can
equally well be written $$\k\Psi = \fr1/{\sqrt2}\k{R}\X\k{D}+
\fr1/{\sqrt2} \k{L}\X\k{\D}, \eq(3.10)$$ since $\k{D}$ and $\k{\D}$
are explicitly $$\k{D} = \fr1/{\sqrt2}\k{H} + \fr1/{\sqrt2}\k{V},$$
$$\k{\D} = \fr1/{\sqrt2}\k{H} - \fr1/{\sqrt2}\k{V}.\eq(3.11) $$ If he
prepares pairs in the state $\k\Psi$, Bob can persuade Alice that he
has given her the ensemble associated with the representation \(3.5)
of $W$ by measuring horizontal-vertical polarization on his own
photons, while he can persuade her that her ensemble is the one
suggested by \(3.6) by measuring diagonal polarization. So if one is
predisposed against quantum nonlocality, one is required to
acknowledge that there is no objective difference between the two
ensembles associated with the representations \(3.5) and \(3.6) of the
density matrix W.  

Einstein was so predisposed: \quote{On one supposition we should, in
my opinion, absolutely hold fast: the real factual situation of the
system $\s_2$ is independent of what is done with the system $\s_1$,
which is spatially separated from the former.\fn} \noindent His
eminently reasonable supposition, has received a bad press because of
what Einstein did with it.  He combined it with his strong intuition
about what constituted a real factual situation, to conclude that
quantum mechanics offers an incomplete description of physical
reality.  Now that John Bell has shown us that it is impossible to
complete the quantum mechanical description of physical reality along
the lines Einstein seemed to call for, we ought to explore the
converse approach to saving his supposition: assume that quantum
mechanics {\it does\/} provide a complete description of physical
reality, {\it insist on Einstein--locality,\/} and see how this
constrains what can be considered {\it physically real\/}.

This gives a rule of thumb for when an internal property
of an isolated physical system can be certified to be {\it objectively
real\/}: a necessary condition for an individual system to have an
objective internal property is that the property cannot change in
immediate response to what is done to a second far-away system that may
be correlated but does not interact with the first.

With this view the reduced density matrix $W$ of a subsystem of a
correlated system can be objectively real, but there can be no
objectively real differences among {\it any\/} of the different
interpretations\ft{It will have (infinitely) many, unless the system
is in a pure state.} of $W$ in terms of ensembles of systems in
different (not-necessarily orthogonal) pure states, associated with
representations of the form $$ W = \sum p_\mu\k{\phi_\mu}\b{\phi_\mu}.
\eq(density)$$ This is the content of a theorem of Gisin, Hughston,
Jozsa, and Wootters,\fn$^,$\fn\ which makes a powerful case for the
fundamental nature of mixed states.

The GHJW theorem establishes that a system (Alice's) can always be
correlated with another system (Bob's) in such a way that when Alice
and Bob have an ensemble of such identically correlated
non-interacting pairs of systems and Alice specifies {\it any\/}
particular representation
\(density) of the density matrix of her system, then Bob can measure
an appropriate observable on his, the result of which will enable him
to instruct Alice how to identify fractions $p_\mu$ of her systems
which have internal properties appropriate to the state $\k{\phi_\mu}$.
Thus Bob can certify as the ``actual'' ensemble any possible
realization of Alice's ensemble in the form of a collection of pure
states $\k{\phi_\mu}$ with probabilities $p_\mu$.  He does this by
giving Alice the information necessary to sort out her collection of
subsystems into any one of those realizations.  In the Appendix I
provide this important fact with an elementary proof, accessible to
senior citizens like Danny and me.

If you want to take Einstein locality seriously, then the GHJW theorem
requires you to acknowledge that there can be no more objective
reality to {\it any\/} of the different possible realizations of a
density matrix, then there is to the different possible ways of
expanding a pure state in terms of different complete orthonormal
sets.  The density matrix of a system must in general be viewed as a
fundamental and irreducible objective property of that system,
containing all information about its internal properties, whether or
not the state the density matrix describes is mixed or pure.

\bigskip
\noindent{{\bf 3. QUANTUM NONLOCALITY: A FABLE}}
\medskip

Here is a more striking version of the ``good-news bad-news'' view of
nonlocality given in Section 2.\ft{What follows is based on
unpublished lecture notes for a talk I gave at Columbia University in
March, 1996.  A similar story, with a somewhat different moral, is told
by Kern and Zeilinger.\fn} This time Carol provides Alice and Bob with
an ensemble of labeled pairs of spin-$\fr1/2$ particles.  Alice gets
one member of each pair and Bob, the other.  Then they each measure
the spin of their own member of pair \#1 along a common direction $\nv_1$,
pair
\#2, along a common direction $\nv_2$, etc, where in each case they
randomly select $\nv_i$ from one of the three orthogonal directions
\xv, \yv, and \zv.  They find that regardless of their choice of
directions, each spin behaves randomly and independently of the the
other.  Is it possible to discriminate between the following
two\ft{One can complicate the tale by adding four more options in
which up and down in cases (I) and (II) refer to the $x$- or
$y$-directions, but I prefer to break the rotational symmetry
and keep the story simple.} explanations for such behavior?

(I) The ensemble Carol gave them consisted of an equally weighted
mixture of spins in the four completely uncorrelated states (up and down are
with respect to the $z$-direction, $a$ and $b$ identify Alice's spin
and Bob's)
$$\k{\ua\ub}, \ \k{\ua\db},\ \k{\da\ub}, \ \k{\da\db}.\eq(1)$$

(II) The ensemble Carol sent was an equally weighted mixture of pairs
in the four states consisting of the singlet state $$\k{0,0}_{ab} =
\c(\k{\ua\db}-\k{\da\ub}),\eq(2) $$ and the three triplet states,
$$\eqalign{ \k {1,1}_{ab} &= \k{\ua\ub},\cr
            \k{1,\oo}_{ab} &= \k{\da\db},\cr \k{1,0}_{ab} & =
	    \c(\k{\ua\db}+\k{\da\ub}).}\eq(3)$$ Two of these four
	    states are highly
correlated --- indeed, they are maximally entangled.  

Note that in either case the global density matrix for the pairs of
Alice and Bob is just the four-dimensional unit matrix.

Zygmund and Yvonne have a debate about whether there is any objective
difference between Case (I) and Case (II).  Are the ensembles of
Alice-Bob pairs objectively different in the two cases?

{\it Zygmund:} Both ensembles are described by exactly the same
two-spin density matrix.  This means there is no way Alice and Bob can
tell the difference.  Therefore there is no difference.

{\it Yvonne:}  You've overlooked the fact that Carol can help Alice
and Bob tell the difference.  If it was Case (I) then Carol can tell
them what their individual results were in every single one of the
cases (about a third of them) in which they happened to have picked the
$z$-direction along which to make their measurements.  She could not
possibly do this in Case (II).  But if the ensemble was Case (II) then
Carol knows which of the pairs were singlets, so she can identify for
Alice and Bob a subset (about a quarter) of their data for which all
their measurements are perfectly anticorrelated, regardless of whether
they were along {\bf x}, {\bf y}, or {\bf z}.  She could not possibly
do this in Case (I).  So in each case she can identify a subset of
Alice and Bob's ensemble of pairs with properties that uniquely
identify that case.  
				   
{\it Zygmund:} But in either case, the collection of data Alice and Bob
accumulate for the runs in which they both happen to pick {\bf z} is
{\it completely random.\/} Carol's trick supposedly demonstrating that
Alice and Bob are dealing with a Case (I) ensemble has nothing to do
with the objective character of what happens in Alice's and Bob's
experiments.  It consists merely in her being able to sort out,
without actually having seen it, the $zz$ portion of their random data
into the four (about equally populated) subsets associated with the
four possible outcomes.

Furthermore, in either case about half the data collected by Alice and
Bob will be perfectly anti-correlated {\it by sheer chance.\/} Carol's
trick supposedly demonstrating that Alice and Bob are dealing with a
Case (II) ensemble consists only in her being able to identify about
half of those perfectly anti-correlated pairs without having access to
their actual data.  The {\it existence\/} of perfect anti-correlations
for {\it some\/} subset of Alice and Bob's data, is again an objective
feature of that data, present in both cases.  

Since the data Carol needs to convince Alice and Bob that their
ensemble is either Case (I) or Case (II) is already there in either
case, all you have demonstrated is that Carol's {\it relation\/} to
the Alice-Bob ensemble --- the kinds of information she can have about
it --- can take various forms.  You have not shown that there is any
objective difference in the two forms the ensemble itself might
have.

{\it Yvonne:}  You can't just talk abstractly about
``Carol's relation'' to the Alice-Bob ensemble.  The fact is that in
preparing the ensemble for Alice and Bob she would actually have had
to do entirely different things to the spins, depending on which
ensemble she prepared.  It's only because Carol didn't provide Alice
and Bob with enough information about what she actually did, that
Alice and Bob are required to describe their ensemble by a density
matrix which is the same in both cases.  In Case (I) Carol had to
prepare pairs of spins with definite values of their z-components,
while in Case II she had to prepare pairs of spins with definite total
angular momentum (and, in the triplet case, with the appropriate
$z$-components of total spin.)  The pairs of spins Alice and Bob possess
had objectively different past histories so it clearly makes sense to
characterize them as objectively different.  

{\it Zygmund:} On the contrary, Carol can prepare Alice and
Bob's ensemble in a way that makes it obvious that Case (I) and Case
(II) are not objectively different.  She can produce each pair she
sends to Alice and Bob by taking {\it two\/} singlet pairs, sending a
member of one pair to Alice, a member of the other pair to Bob, and
keeping the remaining spins herself, labeled so she knows which had
its partner sent to Alice and which to Bob, as well as which Alice-Bob
pair each of her own pairs is associated with.  

If Carol sends Alice and Bob their ensemble of pairs in this way, then
each quartet of spins is in the (rotationally invariant) state
$$\k{\Psi} = \k{{\rm 0,0}}_{ca}\x
           \k{{\rm 0,0}}_{cb} =
           \c\Bigl(\k{\uc1\da} - \k{\dc1\ua}\Bigr)\x
           \c\Bigl(\k{\uc2\db} - \k{\dc2\ub}\Bigr).\eq(4)$$ Carol
can build up the Alice-Bob ensemble by doing this many times.  Then
she can do one of two things: Either (I) she measures the individual
$z$-spins of all the pairs she kept or (II) she measures the total
spin and total $z$-component of each of her pairs of spins. In the
first case she acquires the information necessary to persuade Alice and
Bob that their ensemble is of type (I), since the state $\k{\Psi}$ has
the expansion $$\k{\Psi} = \fr1/2\Bigl(\k{\uc1\uc2}\x\k{\da\db} +
\k{\dc1\dc2}\x\k{\ua\ub} - \k{\uc1\dc2}\x\k{\da\ub} -
\k{\dc1\uc2}\x\k{\ua\db}\Bigr). \eq(5)$$ But in the second case she
acquires the information needed to persuade Alice and Bob that their
ensemble is of type (II), since the state $\k{\Psi}$ also has the
expansion $$\k{\Psi} = \fr1/2\Bigl(\k{1,1}_{cc}\x\k{1,\oo}_{ab} +
                         \k{1,\oo}_{cc}\x\k{1,1}_{ab} -
                         \k{1,0}_{cc}\x\k{1,0}_{ab} +
                         \k{0,0}_{cc}\x\k{0,0}_{ab}\Bigr).\eq(6)$$ 
Since Carol can decide which properties to measure long after Alice and
Bob are in full possession of all their pairs, the two ensembles
cannot be objectively different.  

{\it Yyvonne:} A lovely trick on Carol's part, to be sure, but you
have entirely overlooked the phenomenon of quantum nonlocality.
Carol's measurements actually do change the objective character of
Alice's and Bob's pairs by the {\it SF\/}-process: {\it spukhafte
Fernwirkungen.}

{\it Zygmund:} Even if I grant you the existence of such nonlocal
action, Carol can easily evade your claim that what she does spookily
alters the character of Alice's and Bob's pairs.  Her action need be
taken only {\it after\/} every pair in the Alice-Bob ensemble has been
measured and the data entered in Alice's and Bob's notebooks.  Clearly
if Carol does nothing until Alice and Bob have already collected and
recorded their data, nothing she does can alter the character of the
Alice-Bob ensemble from which those data were extracted.

The only difference between the two cases is that in Case (I) Carol
has collected the information she needs to tell Alice and Bob the
results in the third of their data coming from their $z$-spin
measurements, thereby sacrificing the possibility of getting the
information required to identify to them a quarter of their data that
contains perfect anti-correlations.  In Case (II), on the other hand,
she has collected the information enabling her to identify an
anti-correlated quarter, while sacrificing the possibility of
identifying for them the results of their individual $z$-spin
measurements.  To maintain that the two ensembles are objectively
different you have to maintain that Carol's measurement alters the
objective character of the Alice-Bob pairs in a manner that is not only
non-local in space but backwards in time.

{\it Yvonne:} There is no need for anything to go backwards in
time. If Carol does not make her measurement until long after Alice
and Bob have collected their data then the non-local {\it SF\/} action
has gone forward in time.  But it has gone not from Carol to Alice's
and Bob's ensemble, but from Alice and Bob to the ensemble of pairs
kept by Carol.

As Alice and Bob measure the individual spins of each pair in their
ensemble {\it SF\/} converts the corresponding pair in Carol's
possession into a pair in a product of eigenstates strictly correlated
through \(5) with the results they found for that pair.  (Because
$\k\Psi$ is rotationally invariant it has the same structure \(5)
whether the spin quantization axis is taken to be \zv, \xv, or \yv.)
For Carol later to do her Case (I) trick it is only necessary for her
to measure the $z$-components of the spins of each of her pairs.  The
results will agree with what Alice and Bob found in that third of
their runs in which they measured along \zv.  On the other hand, for
her to do her Case (II) trick she need only identify about a quarter
of her pairs whose spin eigenvalues are opposite, which will enable
aher to identify a quarter of Alice's and Bob's data that is perfectly
anticorrelated.

{\it Zygmund:} But how can she do the Case (II) trick when she doesn't
know the direction along which her spins are polarized, since she
doesn't know whether Alice or Bob measured the spins of any particular
pair along {\bf x}, {\bf y}, or {\bf z}?  To ask them would surely
arouse their suspicions!

{\it Yvonne:} No problem.  Every one of Carol's pairs (about half of
them) that are polarized the {\it same\/} way along their unknown
direction (as a spooky consequence of Alice's and Bob's earlier
measurements) are in a state orthogonal to the singlet state (whatever
that unknown direction may be).  So if Carol measures the total spin
of her pairs and rejects those --- about 3/4 of them --- that have
total spin one, she will be able to identify about half of the
Alice-Bob pairs for which they measured opposite spins.  This provides
her with just the 25\% of perfectly anticorrelated pairs she needs to
fool Alice and Bob into thinking that she had given them an ensemble
of type (II).

{\it Zygmund:} Clever of Carol, to be sure.  But all this shows is
that even in your own terms, the {\it spukhafte Ferwirkungen\/} has
nothing to do with the objective difference between ensembles I and
II.  It has only to do with how Alice and Bob inadvertently telegraph
to Carol the information that enables her to fool them into thinking
that she prepared an ensemble of one or the other type.  

The fact is that the objective character of Alice and Bob's ensemble is
not altered from afar by Carol.  Nor do Alice and Bob spookily send
Carol information about their ensemble.  The fact that the same
actions by Alice, Bob, and Carol can either be interpreted as Carol
altering the character of Alice and Bob's ensemble from afar, or as
Alice and Bob inadvertently transmitting to Carol the information
necessary to enable her ingeniously and deceptively to persuade them of a
particular objective character of their ensemble, illuminates the
fundamental absurdity of either of these two quite different
interpretations.

All that is going on here is that because Carol's pairs of spins are
strongly correlated with the pairs in Alice and Bob's ensemble, Carol
has mutually exclusive options of extracting from her spins various
kinds of information about Alice and Bob's ensemble.  She has the
option of identifying a quarter of their pairs for which they found
opposite spins.  But she also has the option of identifying the
specific results they found in all their $zz$ runs.  

{\it Yvonne\/} [slinking off but muttering under her breath: {\it
Sie wirken doch fern\/}$\,\ldots\,$and what about Bell's theorem$\ldots$?

\bigskip
\noindent{{\bf 4.  INCONCLUSIONS:}$\,$\ft{As a
65th birthday present I dedicate to Danny this new word. Inconclusions are to
conclusions as inconclusive is to conclusive.} {\bf FASHION AT A DISTANCE?}}
\medskip What, indeed, about Bell's theorem?  I am not prepared to
spell out here how Alice's trick in Section II or Carol's in Section
III can be reconciled with the absence of spooky action at a distance
{\it and\/} the completeness of the quantum mechanical description of
physical reality.  I believe the reconciliation is to be found in a
view of physical reality as comprised only of correlations between
different parts of the physical world, and not at all of unconditional
properties possessed by those parts.  In that case the straight answer
to the question posed in my title, like the straight answer to Danny's
inspired but lunatic question in Florence, is ``everything''.  And the
straight question --- to which the straight answer is indeed
``nothing'' --- is ``What do these correlations know about individual
properties?''\ft{I have expanded on such ideas under the noncommittal
rubric of ``the Ithaca interpretation of quantum
mechanics.''\fn$^,$\fn.}

Underlying such a view will be a stronger sense than we currently
possess, of the absurdity of trying to reconcile our perceived
particularity of what actually happens under actual conditions, with
an imagined particularity of what might have happened under conditions
that might have held, but did not.  My guess is that Bohr's not
entirely transparent views on such matters\fn\ might be elucidated\fn\
by rephrasing them in a broader context where the correlated systems
are not, as they invariably are for Bohr, limited to a microscopic
specimen and a classical apparatus.  Such a broader view might reveal
that the problem of ``nonlocality'' and ``the measurement problem''
both stem from the same weakness in current preconceptions about what
constitutes physical reality.  I hope to have more to say about this
(or to be disabused of my current sense of partial illumination) by
Danny's 75th.

All I have tried to do here (aside from offering up a proof of the
GHJW theorem that is accessible to those of us over 60 who don't
understand explanations cast in terms of POVM's) is to suggest that
people may have become too facile in their readiness to blame
everything on (or credit everything to) ``quantum nonlocality.''
Nonlocality seems to me to offer ``too cheap'' a way out of some
deep conundra (to appropriate Einstein's remark to Born about Bohm
theory).  If you push hard on it you can force ``nonlocality'' into
offering some explanations that strike me as just plain silly.

When Abner Shimony first came up with the term ``passion at a
distance'' to characterize the spooky actions at a distance required
by the notion of quantum nonlocality, I thought that by replacing
``action'' with ``passion'' he was emphasizing that the action was on
the interpretive instincts of people (who can experience passion), and
not among the correlated subsystems themselves (which cannot).  But
this ``passion'' has since been taken by just about everybody
(including, I believe, Abner himself) to signify a weaker, spookier
kind of action of the physical systems on each other.  

I would like to suggest that people have been a little too quick in
talking themselves into this widely held position.  To salvage what I
had thought was a very good suggestion of Abner, and as an act of
homage to Danny Greenberger, without the tutorial example of whose
many (superior) bon mots I could never have had the inspiration, I
would like to suggest that the time has come to consider the
possibility that quantum nonlocality is nothing more than {\it fashion
at a distance.\/}\ft{The acronym has not escaped my attention.}
\bigskip
\noindent{{\bf ACKNOWLEDGMENT}}
\medskip Work supported by the National Science Foundation, Grant
No.~PHY9722065.

\bigskip
\noindent{{\bf APPENDIX}}
\medskip
Let Alice have a
system with a density matrix $W$ having a variety of different
expressions in terms of projection operators on finite or countably
infinite collections of (not necessarily orthogonal) pure states.
Each such expression describes Alice's system as if it were in a pure
state $\k{\phi_\mu}$ with probability $p_\mu$:  $$ W = \sum
p_\mu\k{\phi_\mu}\b{\phi_\mu}. \eq(density1)$$ The GHJW theorem
establishes that for any such collection of alternative
interpretations of $W$ it is possible to provide Bob with a system of
his own for which the joint Alice-Bob system has a pure state $\k\Psi$
with the following properties: 

($i$) $W$ is the reduced density matrix for Alice's subsystem. 

($ii$) Associated with {\it any\/} interpretation
\(density1) of Alice's density matrix $W$, there is a corresponding
observable $B$ of Bob's subsystem whose measurement in the state
$\k\Psi$ gives a result $b_\mu$ with probability $p_\mu$.  If Bob
measures $B$ and gets the result $b_\mu$ he is able correctly to
inform Alice that the result of any measurement she makes on her own
system will be as if her system were in the state
$\k{\phi_\mu}$.\ft{In less guarded and clumsy but decidedly more
spooky language, if Bob measures $B$ and gets the result $b_\mu$ then
Alice's system collapses into the state $\k{\phi_\mu}$.} 

The proof that follows is Chris Fuchs's simplification\fn\ of John
Preskill's simplifica\-tion\fn\ of my simplification$^{(8)}$ of the
proof of HJW$^{(6)}$.

The crucial thing to notice is that any two states $\k{\Psi}$ and
$\k{\Psi'}$ of a joint Alice-Bob system leading to the same reduced
density matrix for Alice, $$W =
\tr_{_{{\rm Bob}}}\k\Psi\b\Psi, \eq(reduced)$$ can be related by a
unitary transformation $$ V = 1 \x U\eq(unitary)$$ that acts
non-trivially only on Bob's subspace.  This follows from expanding the
density matrix $W$ of Alice's subsystem in terms of a complete
orthonormal set of eigenvectors:\ft{Some of the $w_i$ can be zero or
the same, so the complete set is not uniquely determined by $W$.  Any
one will do.} $$W = \sum w_i
\k{\al_i}\b{\al_i}.\eq(expand)$$ Because the $\k{\al_i}$ are
complete in Alice's subsystem, an arbitrary state $\k\Psi$ of the
Alice-Bob system has an expansion of the form $$\k{\Psi} =
\sum\k{\al_i}\x\k{\beta_i} \eq(composite)$$ (where the $\k{\beta_i}$ are
in general neither normalized nor orthogonal). The general reduced
density matrix \(reduced) associated with the state \(composite)
can thus be put in the form $$W = \sum
\ip{\beta_j}{\beta_i}\,\,\k{\al_i}\b{\al_j}.\eq(alphas)$$ But if $W$ is of
the special form \(expand), then since the $\k{\al_i}$ {\it are\/} an
orthonormal set, the $\k{\beta_i}$ appearing in
\(composite) are constrained to satisfy $$\ip{\beta_j}{\beta_i} =
w_i\delta_{ij}.\eq(agree)$$ Therefore the $\k{\beta_i}$ can only be non-zero if
$w_i \neq 0$, and when $w_i \neq 0$ they are of the form $$\k{\beta_i} =
\sqrt{w_i}\,\,\k{\gamma_i}\eq(beta)$$ where the $\k{\gamma_i}$ for
$w_i \neq 0$ are orthonormal and can be extended (arbitrarily) to a
complete orthonormal set.  Hence any $\k{\Psi}$ yielding the reduced
density matrix \(reduced) must be of the form $$\k\Psi =
\sum
\sqrt{w_i}\,\,\k{\al_i}\x\k{\gamma_i}.\eq(form)$$ Since the
$\k{\gamma_i}$ are a subset of a complete orthonormal set, all such
forms \(form) do indeed differ only by a unitary transformation of the form 
\(unitary). 

Now consider that representation \(density1) of Alice's density matrix
$W$ containing the largest (possibly infinite) number of distinct
states $\k{\phi_\mu}$. Let Bob's subsystem have a dimension at least
as large as that number, so that it possesses an orthonormal set of
states $\k{\psi_\mu}$ that can be put in one-to-one correspondence
with Alice's states $\k{\phi_\mu}$.  If the composite Alice-Bob system
is in the state $$\k{\Psi} = 
\sum\sqrt{p_\mu}\,\,\k{\phi_\mu}\x\k{\psi_\mu}, \eq(Psi)$$ then Bob
can acquire the information necessary to identify which of the states
$\k{\phi_\mu}$ characterizes Alice's subsystem, by measuring a
nondegenerate observable $B$ of the form $$B = \sum b_\mu
\k{\psi_\mu}\b{\psi_\mu}.\eq(obsB)$$  

Consider now any other
expansion of Alice's density matrix, $$ W =
\sum p'_\mu\k{\phi'_\mu}\b{\phi'_\mu},
\eq(density')$$
Bob can do the same trick using a different state
$$\k{\Psi'} =
\sum\sqrt{p'_\mu}\,\,\k{\phi'_\mu}\x\k{\psi_\mu} \eq(Psi')$$ for the
Alice-Bob system. But since $\k{\Psi}$ and $\k{\Psi'}$ yield the same
reduced density matrix $W$ they must be related by $$\k{\Psi} = 1\x U
\k{\Psi'}.\eq(1xV)$$ Applying $1\x U$ to \(Psi') we learn that the
original state \(Psi) of the Alice-Bob system has the alternative
expansion $$\k{\Psi} =
\sum\sqrt{p'_\mu}\,\,\k{\phi'_\mu}\x\k{\psi'_\mu}, \eq(Psi)$$ where
$$ \k{\psi'_\mu} = U \k{\psi_\mu}.
\eq(psi')$$ Therefore with the composite system in the original state
$\k{\Psi}$, Bob need only measure an observable of the form
$$B' = \sum b_\mu
\k{\psi'_\mu}\b{\psi'_\mu} \eq(obsB)$$ to acquire the information
necessary to persuade Alice that her system is in the state states
$\k{\phi'_\mu}$ with probability $p_\mu'$.

\bigskip
\parindent=0pt
\parskip = 5pt
{\bf REFERENCES.}

1. N. David
Mermin, Phys.~Rev.~Lett.~{\bf 65}, 2272 (1990). 

2. Rudolf Peierls, {\it Surprises in Theoretical Physics},
Princeton Univesity Press, 1979, pps.26-29.

3.  Albert Einstein, {\it Briefwechsel 1916-1955 von Albert Einstein
und Hedwig und Max Born, Kommentiert von Max Born}, Nymphenburger
Verlagshandlung, M\"unchen, 1969.

4. Albert Einstein, {\it Albert Einstein:  Philosopher-Scientist\/},
ed. P. A. Schillp, Open Court, La Salle, Illinois, 1970, p.~85.

5. N.  Gisin, Helv.  Phys. Acta {\bf 62}, 363 (1989).

6.  L.  P. Hughston, R.  Jozsa, and W. K.  Wootters, Phys. Lett. A
{\bf 183}, 14 (1993).

7.  G\"unther Krenn and Anton Zeilinger, Phys.~Rev.~A {\bf 54},
1793-1797 (1996).

8. N. David Mermin, ``The Ithaca Interpretation of Quantum
Mechanics'', Pramana (to be published).  See also Los Alamos e-print
archive, xxx.lanl.gov, quant-ph 9609013.

9. N. David Mermin, ``What is Quantum Mechanics Trying to Tell Us?'',
   Am.~J.~Phys. (to be published). See also quant-ph 9801057.

10. Niels Bohr, Nature {\bf 136}, 65 (1935); Phys.~Rev.~{\bf 48}, 696
(1935). 

11. N. David Mermin, ``Nonlocal Character of Quantum Theory?'',
Am.~J.~Phys. (to be published).

12.  Christopher Fuchs, private communication.

13.  John Preskill, Caltech lecture notes.

\bye